# Content-Based Medical Image Retrieval with Opponent Class Adaptive Margin Loss

Şaban Öztürk, Emin Çelik, Tolga Çukur, Senior Member, IEEE

**Abstract—** Broadspread use of medical imaging devices with digital storage has paved the way for curation of substantial data repositories. Fast access to image samples with similar appearance to suspected cases can help establish a consulting system for healthcare professionals, and improve diagnostic procedures while minimizing processing delays. However, manual querying of large data repositories is labor intensive. Content-based image retrieval (CBIR) offers an automated solution based on dense embedding vectors that represent image features to allow quantitative similarity assessments. Triplet learning has emerged as a powerful approach to recover embeddings in CBIR, albeit traditional loss functions ignore the dynamic relationship between opponent image classes. Here, we introduce a triplet-learning method for automated querying of medical image repositories based on a novel Opponent Class Adaptive Margin (*OCAM*) loss. *OCAM* uses a variable margin value that is updated continually during the course of training to maintain optimally discriminative representations. CBIR performance of *OCAM* is compared against state-of-the-art loss functions for representational learning on three public databases (gastrointestinal disease, skin lesion, lung disease). Comprehensive experiments in each application domain demonstrate the superior performance of *OCAM* against baselines.

Manuscript received XX.XX.XXXX; revised XX.XX.XXXX; accepted XX.XX.XXXX. Date of publication XX.XX.XXXX; date of current version XX.XX.XXXX. The work of T. Çukur was supported by TUBA GEBIP 2015, BAGEP 2017 awards. This research is funded by the Scientific and Technological Research Council of Turkey (TÜBİTAK) under grant number 118C543. (Corresponding author: Şaban Öztürk)

Ş. Öztürk is with the Department of Electrical and Electronics Engineering, Amasya University, TR-05001 Amasya, Turkey, also with the Department of Electrical and Electronics Engineering, Bilkent University, TR-06800 Ankara, Turkey, also with the National Magnetic Resonance Research Center, Bilkent University, TR-06800 Ankara, Turkey, (e-mail: saban.ozturk@amasya.edu.tr).
E. Çelik is with the Department of Electrical and Electronics Engineering, Bilkent University, TR-06800 Ankara, Turkey, and also with the Neuroscience Program, Sabuncu Brain Research Center, Bilkent University, TR-06800 Ankara, Turkey (e-mail: emincelikneuro@gmail.com).
T. Çukur is with the Department of Electrical and Electronics Engineering, Bilkent University, TR-06800 Ankara, Turkey, also with the National Magnetic Resonance Research Center, Bilkent University, TR-06800 Ankara, Turkey, and also with the Neuroscience Program, Sabuncu Brain Research Center, Bilkent University, TR-06800 Ankara, Turkey (e-mail: cukur@ee.bilkent.edu.tr).

## I. Introduction

By providing automated access to medical images that are visually similar to a query image, CBIR systems can aid in the solution of clinical problems such as grading of disease progression, diagnosing multiple concurrent diseases, cross-organ disease assessment, and counseling of medical trainees [1]. Automated retrieval is characteristically performed by comparing the visual contents of the query image against candidate images in a large image repository. Earlier methods assessed visual content in terms of hand-crafted local and global image features [2]. Recent methods instead leverage a latent space where each image is represented as a dense embedding vector [3, 4], and similarity between two images is taken as the distance between their embedding vectors. Representational learning of the latent space is commonly performed via a neural network model trained with an embedding objective to assign similar vectors for visually similar images [5, 6]. Consequently, retrieval is performed by recollecting the set of images that are closest to the query image in the latent space.

The embedding objective involves learning representations that become more discernible as the dissimilarity between images grows. Prominent approaches to do this include point-wise, pair-wise, and triplet-wise methods. Point-wise methods process each image separately, so they can be suboptimal in capturing similarity-based information [7-9]. Pair-wise methods are more amenable to similarity assessments as they process images in pairs [10]. Successful results have been reported with pair-wise methods based on contrastive loss [11-14]. However, these methods have relatively limited training efficiency as they only consider pairs of either similar or dissimilar images, and they can produce weakly discernible representations for similar images. To address these limitations, triplet-wise methods process images in sets of three with anchor (*A*), positive (*P*), and negative (*N*) samples. A triplet loss is commonly used to compare the *A-P* distance between images of the same class (i.e., similar images) against the *A-N* distance between images of different classes (i.e., dissimilar images) [15]. This allows triplet-wise methods to improve the discernability of intra-class representations over pair-wise methods.

The traditional triplet loss enforces the difference between *A-P* and *A-N* distances to remain above a constant margin value [16]. Several recent studies have proposed modifications in the loss function to improve performance in triplet-wise learning. These modifications include linear or non-linear weighting of *A-P* and *A-N* distances in the triplet loss [17-19], and the addition of regularization terms based on the *A-P* distance itself to emphasize its contribution [20, 21]. Yet, similar to the traditional formulation, these methods do not place any explicit constraints on the *P-N* distance [15], which can lower the segregation between *P* and *N* samples from opponent classes [22]. Furthermore, these methods pervasively use constant margin values that can elicit suboptimal performance since the ideal margin varies across datasets and across the training iterations for a given dataset.

Here we propose an Opponent Class Adaptive Margin (*OCAM*) loss to improve triplet-wise representational learning in CBIR tasks. Addressing two main limitations of the traditional triplet formulation, *OCAM* incorporates the *P-N* distance to enforce better segregation between opponent classes, and it leverages an adaptive margin determined based on the current segregation between the opponent classes. As such, *OCAM* can learn more discernible embedding vectors for medical imaging that can facilitate subsequent image retrieval. Demonstrations are provided on gastrointestinal, skin and lung images based on Euclidean and Hamming retrieval codes. Our experiments indicate that *OCAM* exhibits superior performance against competing baselines. Our main contributions are summarized below:

- *OCAM* leverages an adaptive margin between *A-P* and *A-N* distances to improve conformity to the image distribution per dataset, without necessitating manual intervention.
- *OCAM* incorporates the *P-N* distance in the embedding objective to enhance the discernability of opponent image classes in the latent space.
- Superior retrieval performance is obtained in various anatomies with the *OCAM*-based CBIR method.

## II. THEORY

### a. CBIR

Given a single query image, CBIR methods aim to retrieve a finite subset of *Z* images from a repository with high similarity in visual content to the query. Let's consider a repository $D=\{X,Y\}^K$ consisting of *K* medical images $X=\{x_1,x_2,\ldots,x_K\} \in R^{d \times K}$, where $x_k \in R^{dx(1 \leq k \leq K)}$ is the *k*th sample of *X*, $Y=\{y_1,y_2,\ldots,y_K\}$ represents image labels and *d* denotes the image dimensionality. For assessing the similarity of visual content, the images are typically mapped onto a latent space via $\Omega: x_k \rightarrow E_k$, where $\Omega$ denotes the projection from the original image space onto the latent space, and the set of embedding vectors for the images are given as $E=\{E_1,E_2,\ldots,E_K\} \in R^S$. To retrieve the most similar images from the repository, a search for the *Z* nearest neighbors (NN) to the query is required [23]:

$$NN(x_q, Z) = \left\{ f(E_q, E_k) \in F : \left| F \cap (-\infty, f(E_q, E_k)) \right| < Z \right\} \quad (1)$$

where $E_q$ denotes the embedding of the query image, *f* is a distance metric, and *F* denotes the set of distances between the embeddings of repository images and the query image. It is possible to conduct the search by ranking images according to the Euclidean distance of their continuous embedding codes [24]. While search based on continuous codes can be more sensitive, it also introduces a computational burden for large repositories. For improved search efficiency, a binary hash code can be generated for each image based on its embedding, $B=\{b_1,b_2,\ldots,b_K\} \in \{-1,+1\}^{S \times K}$ [25] via a binarization operation:

$$b_k = \text{sgn}(E_k) = \begin{cases} +1, & E_k^i \geq 0 \\ -1, & E_k^i < 0 \end{cases}, \quad i=1,2,\ldots,S \quad (2)$$

Images can then be ranked according to the Hamming distances of their hash codes [26]. While hash codes improve efficiency, their retrieval performance is inevitably dependent on the representation power of the codes. Therefore, deep learning (DL) methods are commonly adopted to obtain hash codes with high representation power to capture the underlying dense embedding vectors for medical images.

### b. Learning of Image Embeddings

A powerful approach for capturing latent representations of images relies on triplet-wise learning [27]. The traditional triplet loss (*Triplet*) for representational learning samples a set of three images ($x_A$, $x_P$, and $x_N$) from the repository, as illustrated in Fig. 1. Assuming access to information regarding whether any pair of images belong to the same image class, once $x_A$ is initially selected, $x_P$ is drawn from the same class, whereas $x_N$ is drawn from the opponent class. Embeddings of the images in the triplet are computed via $\Omega$, $\Omega: x_A, x_P, x_N \rightarrow E_A, E_P, E_N$, $E \in R^S$. The loss is then expressed as:

$$Triplet(E_A, E_P, E_N) = \max(0, f(E_A, E_P) - f(E_A, E_N) + \alpha) \quad (3)$$

where *α* represents the margin parameter. A first limitation of the traditional formulation is that, for a random selection of the image triplet, it is possible that $f(E_A,E_P) \geq f(E_P,E_N)$ even if the condition in Eq. (3) is satisfied as $f(E_A,E_P)+\alpha \leq f(E_A,E_N)$. This lack of explicit control over $f(E_P,E_N)$ can in turn lower the discernability of learned embeddings $E_P$, $E_N$. A second limitation is that the margin value *α* requires manual tuning for each dataset, which can be labor intensive. Moreover, the margin value is kept fixed across the entire course of training, so its value can be suboptimal for certain portions of the training process.

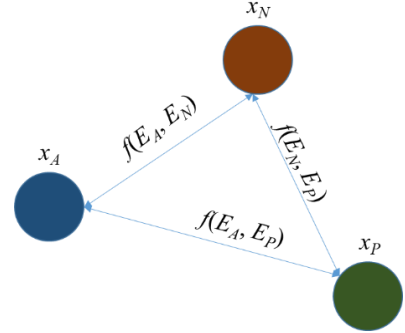

**Figure 1**. To define a triplet loss, three image samples are selected. These include an anchor sample ($x_A$, *in blue color*), a positive sample similar to the anchor ($x_P$, *in green color*), and a negative sample dissimilar to the anchor ($x_N$, *in red color*). Links between image samples illustrate their relative distances.

### c. OCAM

Here we introduce an improved triplet-wise method for medical image retrieval based on a new loss function, *OCAM*. *OCAM* introduces two key technical improvements over the traditional triplet loss. The traditional formulation ignores the *P-N* distance, $f(E_P,E_N)$, so it can yield insufficient segregation between opponent image classes. To address this issue, *OCAM* explicitly incorporates $f(E_P,E_N)$ in the loss function as inspired by a recent computer vision study on person reidentification [22]. Note that there are two samples in each image triplet belonging to the positive class, so weighting the distance terms equally can introduce unwanted biases towards the positive class. To avoid potential biases, a balanced weighting is instead adopted here: $(f(E_A,E_P)-(f(E_A,E_N)+f(E_P,E_N))/2)$. As such, *OCAM* improves inter-class segregation by not only extending the *A-P* distance over the *A-N* distance, but also maintaining a relatively large *P-N* distance.

The traditional formulation also uses a constant margin value to segregate opponent classes. However, a static margin can result in suboptimal performance since $f(E_A,E_P)-f(E_A,E_N)$ inherently changes across the training process. In particular, an *α* value suited for early iterations where $(f(E_A,E_P)-(f(E_A,E_N)+f(E_P,E_N))/2)$ is relatively small will be rendered suboptimal towards later iterations as inter-class segregation increases, slowing down the learning process. To address this limitation, *OCAM* introduces an adaptive margin value inspired by the recent success of adaptive methods in classification tasks [16, 28, 29]. Here we propose to leverage a margin value $\alpha_{adaptive}=(1-f(E_P,E_N))/2$ that is a function of $f(E_P,E_N)$ so as to improve the *P-N* separation, avoiding the need for introducing a user-controlled parameter.

Taken together, these design elements result in the following formulation in *OCAM*:

$$OCAM(E_A, E_P, E_N) = \max\left(0, f(E_A, E_P) - \frac{(f(E_A, E_N) + f(E_P, E_N))}{2} + \frac{(1 - f(E_P, E_N))}{2}\right) \quad (4)$$

which can be further simplified as:

$$OCAM(E_A, E_P, E_N) = \max\left(0, f(E_A, E_P) - \left(\frac{f(E_A, E_N) + 2f(E_P, E_N) - 1}{2}\right)\right) \quad (5)$$

To learn embeddings with high representation capability, a cosine distance measure that ranges in [0,1] is used [12, 13, 25]:

$$f(E_i, E_j) = \frac{(1 - \cos(E_i, E_j))}{2} \quad (6)$$

Following the training of a neural network $\Omega_{Trained}$ according to the loss in Eq. 5, inference can be performed for a query image $x_q$ and a test repository $D_{Test}=\{X_{Test}\}^M$ consisting of *M* test images $X_{Test}=\{x_1,x_2,…,x_M\}\in R^{d\times M}$, where $x_m\in R^{d\times(1\leq m\leq M)}$ is the *m*th sample of $X_{Test}$. Both the query image and test images in the repository are projected into the latent space via $\Omega_{Trained}$. Afterwards, a search is performed to identify the most similar embeddings in $E_{Test}=\{E_1,E_2,…,E_M\}\in R^S$ to $E_q$ as illustrated in Fig. 2.

## III. METHODS

### a. Datasets

We demonstrated the CBIR performance of *OCAM* on three medical image datasets. The KVASIR dataset [30] contained endoscopic images from eight classes of gastrointestinal disease, with 1000 images per class. The ISIC 2019 dataset [31] contained dermoscopic images from eight classes of skin lesions, with the number of images in each class varying from 239 to 12875. The X-RAY dataset [32] contained radiographic images from four classes of lung disease, with the number of images in each class varying from 4497 to 5768. All images were downsampled to 256x256 pixels. 85% of the images were used for model training, and 15% were reserved for model testing.

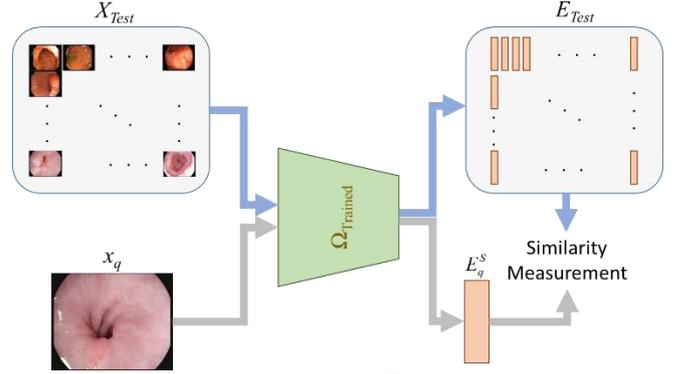

**Figure 2**. For CBIR, given a query image $x_q$, $E_q$ is computed based on a trained network $\Omega_{Trained}$ and compared against the embedding vectors in the repository $E_{Test}$ to retrieve similar images. Candidate images from the repository can be ranked based on the Euclidean distance of continuous embedding codes for the accuracy or based on Hamming distance of binarized hash codes for efficiency.

### b. Architectural Details and Model Implementation

A Siamese neural network model was leveraged to capture the latent representations of medical images with *OCAM* (Fig. 3). The weights of subnetworks processing anchor, positive and negative samples were tied. The subnetworks were designed by considering popular convolutional neural network (CNN) architectures in computer vision: VGG16 [33], ResNet50 [34], InceptionV3 [35], MobileNetV2 [36], DenseNet169 [37], and EfficientNetB3 [38]. To obtain *S*-bit hash codes of image embeddings at the output of the Siamese network, the final fully connected (FC) layers in backbone CNNs were replaced with a dropout layer at 0.3 dropout rate and a dense layer of length *S*. Separate models were designed for *S*=16 and *S*=64.

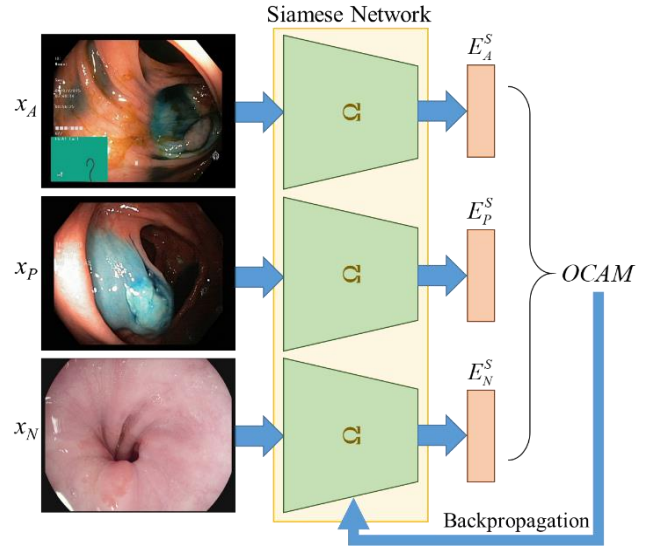

**Figure 3**. Representational learning with *OCAM*. Embedding vectors $E_A$, $E_P$, and $E_N$ for a randomly selected image triplet $x_A$, $x_P$, and $x_N$ are generated with a Siamese network. The Siamese network contains backbone CNNs (Ω) with tied weights across anchor, positive and negative samples. CNN parameters are trained to minimize the *OCAM* loss.

Models were implemented in the TensorFlow framework and executed on an NVidia RTX 3090 GPU. Model training was performed with the Adam optimizer, a batch size of 20 (corresponding to a selection of 60 images per batch due to triplet sampling), and a learning rate of $10^{-5}$. Backbone CNNs were initialized with weights pretrained on ImageNet for object classification. Models were trained until convergence on each dataset, and because the dataset sizes varied, the number of epochs was 4500 for KVASIR, 45000 for ISIC 2019, and 5000 for X-RAY datasets. CBIR performance was separately examined for query search with continuous codes in the Euclidean space of dense embedding vectors [2, 10, 15], and for query search with binary codes in the Hamming space of the embedding vectors [4, 12, 26].

### c. Competing Methods

Comparisons were performed against CBIR methods that perform representational learning via state-of-the-art point-wise, pair-wise, and triplet-wise loss functions. We adhered to the hyperparameters given in the original studies for each competing method described below.

*Neural Codes:* The *Neural Codes* method trains a backbone CNN architecture for a classification task and then discards its output layer to use it in a retrieval task [39]. VGG16 was taken as the backbone CNN, and the dimensionalities of two FC layers prior to softmax were changed to 2048 and $S$. In each dataset, the backbone CNN was trained to detect image classes by minimizing a point-wise categorical cross-entropy loss as in Eq. 7. For retrieval, the softmax layer was removed, and the image embedding was taken as the vector of $S$-dimensional activations prior to the softmax layer in the trained CNN.

$$Neural\ Codes = -\frac{1}{K}\sum_{k=1}^{K}\sum_{j=1}^{J} y_{k,j} \log\left(\Omega\left(x_{k,j};\theta\right)\right) \quad (7)$$

where $K$ represents the total number of samples in the training repository, $J$ represents the number of different classes in the training repository, and $\theta$ is the set of parameters of the $\Omega$.

*Contrastive:* The *Contrastive* method employs a traditional pair-wise contrastive loss on two anchor images $(x_{A1}, x_{A2})$ based on their similarity label ($L$, 0=dissimilar, 1=similar). For this purpose, two images are randomly selected from the repository: $x_{A1}, x_{A2}$. If these two anchor images belong to the same class, parameter $L$ is set to 1, if they belong to different classes, parameter $L$ is set to 0. As such, $E_{A1}$ and $E_{A2}$ can be derived by minimizing the following loss with $\alpha$ taken as 0.2 [10, 40]:

$$Contrastive = L\frac{1}{2}f\left(E_{A1},E_{A2}\right)^2 + (1-L)\frac{1}{2}\max\left(0,\alpha-f\left(E_{A1},E_{A2}\right)\right)^2 \quad (8)$$

*Triplet:* The *Triplet* method employs a traditional triplet loss function with $\alpha$ taken as 0.2 [15, 16, 20]:

$$Triplet\left(E_A,E_P,E_N\right) = \max\left(0, f\left(E_A,E_P\right) - f\left(E_A,E_N\right) + \alpha\right) \quad (9)$$

*TriEP:* The *TriEP* method uses weighting coefficients for distance measures in the triplet loss, and selects the hardest positive and negative samples [17].

$$TriEP\left(E_A,E_P,E_N\right) = \left[\sigma_1\sigma_2 \max\left(f\left(E_A,E_P\right)\right) - \beta_1\beta_2 \min\left(f\left(E_A,E_N\right)\right) + \alpha\right] \quad (10)$$

where $\sigma_1, \sigma_2, \beta_1, \beta_2$ represent the weighting coefficients, and $\alpha=0.3$, $\sigma_1=2.04$, $\sigma_2=1.71$, $\beta_1=0.83$, $\beta_2=0.64$ were used.

*WABT:* The *WABT* method performed triplet-wise learning where the anchor sample is scaled prior to distance calculations to improve performance [18]:

$$WABT\left(E_A,E_P,E_N\right) = \max\left(0, f\left(rE_A,E_P\right) - f\left(rE_A,E_N\right) + \alpha\right) \quad (11)$$

where $r$ denotes the scaling coefficient, and $\alpha=1$ and $r=3$ were used.

*dmTri:* The *dmTri* method uses a dynamic margin value by normalizing the loss function in the traditional triplet formulation with the sum of the *A-P* distance and $\alpha$ [41], with $\alpha$ taken as 0.2:

$$dmTri\left(E_A,E_P,E_N\right) = \max\left(0, 1 - \frac{f\left(E_A,E_N\right)}{f\left(E_A,E_P\right)+\alpha}\right) \quad (12)$$

*CondTri:* The *CondTri* method augments the traditional triplet loss with a weighted regularization term based on the individual distance measures [20], with $\alpha=0.2$ and the regularization weight $\delta=0.1$:

$$CondTri\left(E_A,E_P,E_N\right) = \max\left(0, f\left(E_A,E_P\right) - f\left(E_A,E_N\right) + \alpha\right) + \delta\left[\frac{f\left(E_A,E_P\right) + f\left(E_A,E_N\right)}{2}\right] \quad (13)$$

*CTLL:* The *CTLL* method augments the traditional triplet loss with a weighted and biased regularization term based on the difference norm of the embeddings for anchor and positive samples [21], with $\alpha=1$, the regularization weight $\kappa=0.01$, and the regularization bias $\gamma=0.01$:

$$CTLL\left(E_A,E_P,E_N\right) = \max\left(0, f\left(E_A,E_P\right) - f\left(E_A,E_N\right) + \alpha\right) + \left(\kappa\left(f\left(E_A-E_P\right)\right) - \gamma\right) \quad (14)$$

### d. Performance Metrics

CBIR performance was measured using the precision metric for a total of $Z$ retrieved images (P@Z) and the mean average precision (mAP) metric [10, 14, 25, 42]. Given a repository $D$ with $J$ image classes, P@Z was computed as the across-class average of class-specific (P@Z)$_j$. To compute (P@Z)$_j$, a single test image from the $j$th class was taken as the query, and retrieval was attempted on the entire test set excluding the query image. This process was repeated across all possible query images for the $j$th class. Afterwards, class-specific performance (P@Z)$_j$ and P@Z were computed:

$$\left(P@Z\right)_j = \frac{1}{n_j Z}\sum_{i=1}^{n_j}\sum_{z=1}^{Z} t\left(x_i^j, NN\left(x_i^j,Z\right)_z\right)$$
$$P@Z = \frac{1}{J}\sum_{j=1}^{J}\left(P@Z\right)_j \quad (15)$$

where $n_j$ denotes the number of samples in the $j$th class, $x_i^j$ is the ith query image from the jth class, $NN(.)_z$ denotes the zth element retrieved from the repository among the set of Z images, and $t$ is an indicator function that returns 1 if its inputs are from the same class, and returns 0 otherwise. (mAP)$_j$ and mAP were calculated as follows:

$$（mAP）_j = \frac{1}{n_j Z} \sum_{i=1}^{n_j} \sum_{z=1}^{Z} \frac{t\left(x_i^j, NN\left(x_i^j, Z\right)_z\right)}{z} \quad (16)$$

$$mAP = \frac{1}{J} \sum_{j=1}^{J} (mAP)_j$$

Note that a high P@Z score can be attained when the number of images retrieved from the same class as the query image is high. Meanwhile, mAP does not only consider the raw number of correctly retrieved images but it also reflects the order in which images are retrieved. In particular, a high mAP score can be attained when images from the same class as the query are retrieved in higher ranks among Z images compared to images from other classes.

## IV. RESULTS

### a. Ablation Studies

Several ablation studies were performed to demonstrate the value of the individual components in *OCAM*, including the backbone CNN, the *P-N* distance loss component, and the adaptive margin value. First, retrieval performance was evaluated for *OCAM* variants based on six different backbone CNNs. mAP performances with *S*=64 are given in Table I (E represents Embedding space and H represents Hamming space). The variants with VGG16 yield optimal or near-optimal performance across all CBIR tasks, so VGG16 was designated as the backbone CNN in the remaining experiments.

The retrieval performance of *OCAM* was also evaluated comparatively against a variant ablated of the *P-N* distance $f(E_P,E_N)$ (w/o $f(E_P,E_N)$), a variant ablated of the adaptive margin $\alpha_{adaptive}$ (w/o $\alpha_{adaptive}$), and a variant ablated of both the *P-N* distance and the adaptive margin (w/o $f(E_P,E_N)$ and $\alpha_{adaptive}$). Performances of variant models are shown in Fig. 4 with *S*=64. Overall, *OCAM* outperforms all ablated variants and the performance benefits grow towards larger Z values. Furthermore, both $f(E_P,E_N)$ and $\alpha_{adaptive}$ elements in *OCAM* contribute to CBIR performance in terms of P@Z and mAP. These observations are seen in both Euclidean and Hamming spaces, although there is an expected performance decline for all methods in Hamming space due to the binarization of embeddings.

TABLE I
RETRIEVAL PERFORMANCE OF *OCAM* ACROSS BACKBONE CNNs

| | KVASIR (*S*=64) | | ISIC 2019 (*S*=64) | | X-RAY (*S*=64) | |
|---|---|---|---|---|---|---|
| | E | H | E | H | E | H |
| VGG16 [31] | **88.74** | **85.93** | 72.95 | **71.48** | **87.32** | **86.41** |
| ResNet50 [32] | 87.03 | 82.22 | 72.44 | 70.91 | 86.52 | 85.71 |
| InceptionV3 [33] | 84.49 | 80.13 | 70.23 | 69.16 | 84.60 | 83.71 |
| MobileNetV2 [34] | 82.13 | 75.62 | 72.16 | 70.76 | 83.57 | 82.66 |
| DenseNet169 [35] | 88.52 | 84.37 | 72.98 | 71.20 | 85.98 | 85.34 |
| EfficientNetB3 [36] | 85.47 | 81.33 | **73.13** | 71.13 | 83.76 | 82.52 |

The main motivation for *OCAM* is to improve the discernability of latent space representations for images from separate classes. We reasoned that if *OCAM* improves representational discernability over traditional triplet loss, then the learned embedding vectors should be better segregated in the embedding space. To test this prediction, we projected the embedding vectors learned using *OCAM* and *Triplet* into a two-dimensional space via t-Distributed Stochastic Neighbor Embedding (t-SNE) [9, 15, 18]. The image projections in the t-SNE space are displayed in Fig. 5. For *OCAM*, images from different classes project to spatially segregated clusters that are well separated from each other. In contrast, for *Triplet*, the separation between clusters is relatively lower, and samples from distinct clusters can occur in spatially proximate locations. These results indicate that *OCAM* improves inter-class discernability over the traditional triplet method.

To assess the influence of this representational discernability on retrieved images, we visually inspected images retrieved based on *OCAM* and based on *Triplet*. Fig. 6 depicts Z=10 images retrieved in response to a query image randomly selected from the test set of each of the three datasets. In general, *OCAM*-retrieved images show higher visual similarity to the query image. Note that CBIR methods can occasionally retrieve image samples from opponent classes. An important performance criterion for CBIR methods is the retrieve rank of opponent-class samples (which is also captured in the mAP metric). The retrieve rank of opponent samples in *OCAM* is lower than that for *Triplet*, indicating that *OCAM* is more resilient against erroneous sample selection.

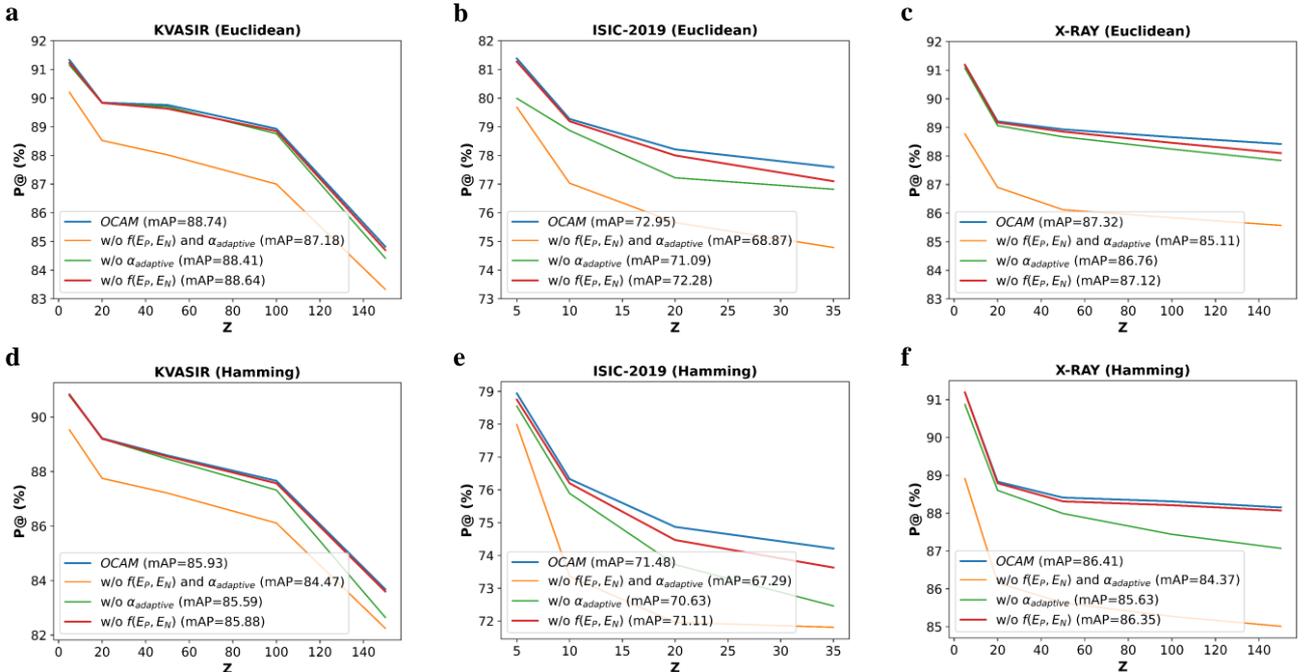

**Figure 4**. Precision performances of OCAM variants were measured based on 64-dimensional embedding vectors. P@Z metrics are plotted for the KVASIR (a,d), ISIC 2019 (b,e), and X-RAY (c,f) datasets across various numbers of retrieved images (*Z*). Results for Euclidean space are given in top row, and those for Hamming space are given in bottom row. mAP metrics for each variant are listed in parantheses.

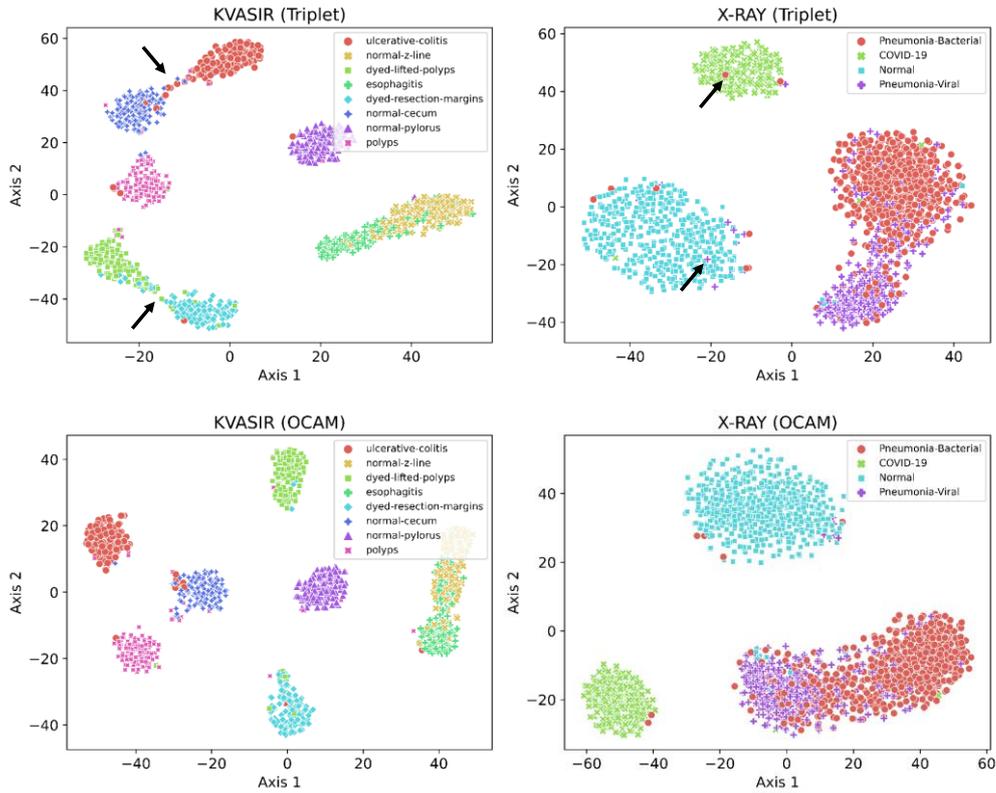

**Figure 5**. The latent space distribution of images were visualized using two-dimensional t-SNE projections of learned embedding vectors. Representative results are displayed for embeddings learned based on the traditional triplet method (*Triplet*) versus *OCAM*. *OCAM* yields more discrete clusters than *Triplet*. Black arrows highlight problematic samples with insufficient inter-class segregation. The distinct classes and their labels are described in respective legends for each dataset.

### b. CBIR Experiments

Next, *OCAM* was comparatively demonstrated for CBIR tasks against state-of-the-art representational learning approaches. Competing methods included a point-wise learning approach (*Neural Codes*) [39], a pair-wise learning approach (*Contrastive*) [10], and several triplet-wise learning approaches including traditional triplet learning (*Triplet*) [15], expansion-pool tri-hard learning (*TriEP*) [17], weighted anchor based triplet learning (*WABT*) [18], dynamic margin triplet learning (*dmTri*) [41], conditional triplet learning (*CondTri*) [20], and constrained triplet loss layer learning (*CTLL*) [21]. Experiments were conducted in the continuous Euclidean space to demonstrate the full performance of learned representations, and in binarized Hamming space to demonstrate the performance in real-time applications that require computational efficiency.

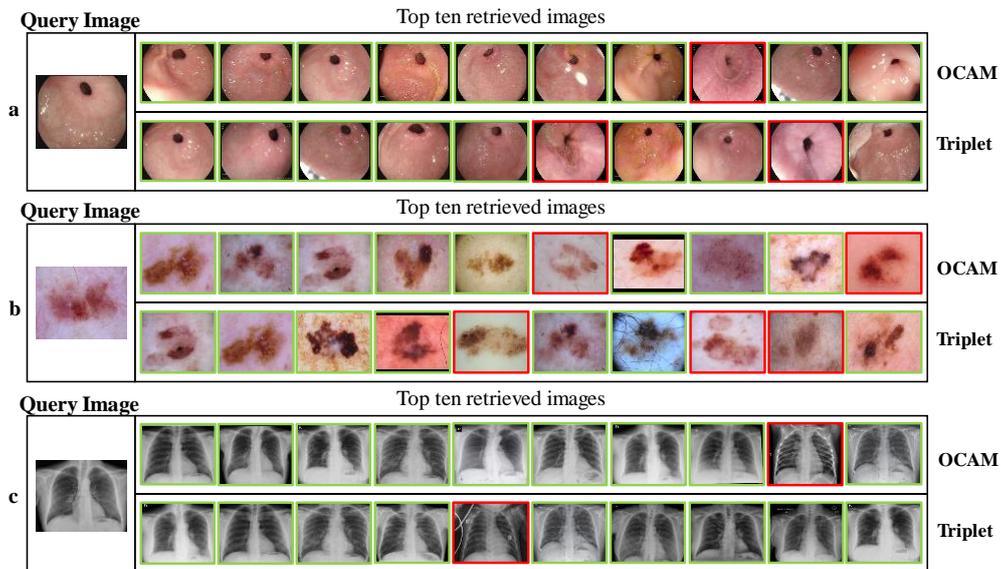

**Figure 6**. Representative retreival results for *Z*=10 images given a single query image. Results are shown for the a) KVASIR, b) ISIC-2019, c) X-RAY datasets. Images retreived from the same class as the query image are marked in green bounding boxes, whereas opponent-class samples are marked in red.

TABLE II
RETRIEVAL PERFORMANCE IN KVASIR DATASET (EUCLIDEAN)

|  | S=16 | | | | | | S=64 | | | | | |
| --- | --- | --- | --- | --- | --- | --- | --- | --- | --- | --- | --- | --- |
|  | P@5 | P@20 | P@50 | P@100 | P@150 | mAP | P@5 | P@20 | P@50 | P@100 | P@150 | mAP |
| *Neural Codes* [39] | 86.22 | 83.98 | 82.28 | 79.81 | 76.12 | 78.07 | 88.21 | 84.61 | 82.62 | 80.97 | 76.48 | 79.36 |
| *Contrastive* [10] | 86.86 | 84.12 | 82.62 | 80.57 | 76.58 | 80.63 | 89.68 | 87.28 | 86.69 | 85.37 | 81.49 | 85.59 |
| *Triplet* [15] | 90.13 | 87.61 | 86.88 | 85.52 | 81.32 | 85.22 | 90.20 | 88.52 | 88.02 | 87.00 | 83.33 | 87.18 |
| *TriEP* [17] | 87.83 | 85.59 | 84.69 | 83.06 | 78.64 | 83.01 | 86.78 | 83.54 | 81.98 | 79.83 | 75.97 | 80.10 |
| *WABT* [18] | 81.45 | 76.67 | 75.23 | 74.26 | 72.17 | 74.80 | 82.45 | 78.28 | 76.32 | 74.96 | 72.40 | 75.71 |
| *dmTri* [41] | 90.47 | 88.35 | 87.38 | 85.97 | 81.90 | 85.73 | 90.32 | 88.40 | 87.56 | 86.16 | 82.01 | 86.27 |
| *CondTri* [20] | 87.48 | 85.58 | 84.81 | 83.26 | 79.75 | 83.63 | 90.96 | 88.84 | 88.13 | 87.22 | 83.77 | 87.30 |
| *CTLL* [21] | 89.72 | 87.07 | 86.47 | 85.24 | 80.89 | 85.09 | 89.88 | 87.89 | 87.39 | 86.68 | 82.87 | 86.71 |
| *OCAM* | 90.75 | 88.97 | 88.64 | 87.40 | 84.13 | **87.30** | 91.33 | 89.84 | 89.76 | 88.93 | 84.82 | **88.74** |

TABLE III
RETRIEVAL PERFORMANCE IN ISIC 2019 DATASET (EUCLIDEAN)

|  | S=16 | | | | | S=64 | | | | |
| --- | --- | --- | --- | --- | --- | --- | --- | --- | --- | --- |
|  | P@5 | P@10 | P@20 | P@35 | mAP | P@5 | P@10 | P@20 | P@35 | mAP |
| *Neural Codes* [39] | 66.76 | 65.18 | 63.44 | 61.97 | 56.15 | 68.97 | 66.61 | 64.43 | 62.76 | 57.90 |
| *Contrastive* [10] | 68.08 | 65.16 | 64.98 | 63.94 | 58.85 | 72.49 | 70.19 | 67.78 | 66.14 | 60.27 |
| *Triplet* [15] | 77.70 | 74.84 | 73.21 | 72.48 | 67.92 | 79.68 | 77.03 | 75.66 | 74.79 | 68.87 |
| *TriEP* [17] | 76.70 | 73.55 | 72.28 | 71.33 | 66.45 | 78.60 | 75.26 | 73.61 | 72.49 | 67.84 |
| *WABT* [18] | 73.44 | 70.35 | 68.76 | 67.78 | 63.64 | 76.68 | 73.59 | 71.99 | 71.10 | 66.67 |
| *dmTri* [41] | 77.75 | 74.81 | 73.25 | 72.39 | 68.03 | 80.26 | 78.09 | 76.92 | 76.15 | 71.27 |
| *CondTri* [20] | 78.08 | 75.16 | 73.98 | 72.94 | 68.74 | 79.71 | 77.24 | 76.22 | 75.48 | 70.75 |
| *CTLL* [21] | 77.75 | 74.75 | 73.19 | 72.21 | 67.50 | 78.64 | 75.74 | 74.32 | 71.18 | 69.15 |
| *OCAM* | 79.29 | 77.00 | 75.79 | 75.20 | **70.77** | 81.38 | 79.27 | 78.21 | 77.59 | **72.95** |

TABLE IV
RETRIEVAL PERFORMANCE IN X-RAY DATASET (EUCLIDEAN)

|  | S=16 | | | | | | S=64 | | | | | |
| --- | --- | --- | --- | --- | --- | --- | --- | --- | --- | --- | --- | --- |
|  | P@5 | P@20 | P@50 | P@100 | P@150 | mAP | P@5 | P@20 | P@50 | P@100 | P@150 | mAP |
| *Neural Codes* [39] | 86.85 | 84.27 | 83.81 | 83.54 | 81.66 | 81.17 | 87.09 | 84.88 | 84.22 | 84.16 | 82.71 | 81.86 |
| *Contrastive* [10] | 88.06 | 85.62 | 85.22 | 85.03 | 84.67 | 84.40 | 88.09 | 85.88 | 85.46 | 85.11 | 84.89 | 84.72 |
| *Triplet* [15] | 88.60 | 86.13 | 85.53 | 85.41 | 85.21 | 84.48 | 88.77 | 86.90 | 86.12 | 85.84 | 85.57 | 85.11 |
| *TriEP* [17] | 88.10 | 85.45 | 84.82 | 84.77 | 84.63 | 84.06 | 88.32 | 86.20 | 85.70 | 85.42 | 85.18 | 84.68 |
| *WABT* [18] | 89.28 | 87.39 | 86.85 | 86.58 | 86.31 | 85.41 | 87.47 | 85.40 | 85.22 | 85.18 | 85.02 | 84.68 |
| *dmTri* [41] | 88.65 | 86.03 | 85.37 | 85.09 | 84.94 | 84.12 | 88.36 | 86.14 | 85.78 | 85.64 | 85.40 | 84.82 |
| *CondTri* [20] | 88.78 | 86.63 | 86.02 | 85.67 | 85.48 | 84.63 | 88.74 | 86.45 | 85.82 | 85.41 | 85.16 | 84.63 |
| *CTLL* [21] | 88.32 | 85.78 | 85.07 | 84.84 | 84.69 | 84.07 | 88.53 | 86.25 | 85.96 | 85.72 | 85.47 | 84.88 |
| *OCAM* | 90.42 | 88.49 | 87.84 | 87.53 | 87.20 | **85.66** | 91.19 | 89.21 | 88.93 | 88.66 | 88.42 | **87.32** |

*Euclidean Space:* Tables II-IV list retrieval performances of competing methods on the test sets of the KVASIR, ISIC 2019, and X-RAY datasets, respectively. Note that the number of samples included in the minority class varied across datasets, which places a practical limit on the number of images that can be retrieved. Accordingly, retrieval was performed for up to Z=150 images in KVASIR and X-RAY, and up to Z=35 images in ISIC 2019 that had a more dramatic class imbalance. Compared to the second-best method, *OCAM* improves mAP by 1.57% at *S*=16 and 1.44% at *S*=64 in KVASIR, by 2.03% at *S*=16 and 3.80% at *S*=64 in ISIC 2019, and by 0.25% at *S*=16 and 2.21% at *S*=64 in X-RAY.

*Hamming Space:* Tables V-VII list retrieval performances of competing methods on the test sets of the KVASIR, ISIC 2019, and X-RAY datasets, respectively. As expected, performances of all methods are moderately lower in the Hamming versus Euclidean space due to loss of information during the binarization of embedding vectors. In general, *OCAM* outperforms competing methods in CBIR performance. Compared to the second-best method, *OCAM* improves mAP by 3.85% at *S*=16 and 1.46% at *S*=64 in KVASIR, by 1.07% at *S*=16 and 2.71% at *S*=64 in ISIC 2019, and by 2.01% at *S*=64 in X-RAY. Meanwhile, it yields a moderately lower mAP by 0.22% compared to *WABT* with Hamming codes at *S*=16 in X-RAY.

Overall, we observe that the performance benefits of OCAM against competing methods are relatively larger in the ISIC 2019 dataset versus the KVASIR and XRAY datasets, where performance metrics are generally higher. Note that the difficulties of the representational learning task and subsequent image retrieval task grow with higher class imbalances in the repository. Thus, this finding suggests that OCAM shows improved reliability against class imbalances compared to baselines.

## V. DISCUSSION

CBIR methods promise fast, automated access to images from a medical repository that are visually similar to a query image to facilitate downstream assessments with visual examples. Current CBIR approaches typically rely on the learning of embedding vectors with high representation capability for visual image features, as subsequent retrieval performance depends critically on the representational quality of the embedding vectors [17, 25, 42-44]. Here, we introduced a novel triplet-wise representational learning method, *OCAM*, for improved CBIR performance on medical image repositories. OCAM leverages triplet learning with an improved objective that considers distances between positive and negative classes and an adaptive margin value. Explicit consideration of *f(P,N)* enables *OCAM* to improve inter-class segregation in the embedding space, while the adaptive margin value improves performance by automatically tuning the margin value depending on the learned representations at each stage of the training process.

CBIR performance can be influenced by the code length and the code type used to perform the image similarity assessments. In our experiments, CBIR was performed based on two different code lengths, *S*=16 and 64. As would be expected, we observed that retrieval performances increased with higher *S*. On average, the increase in mAP performance of *OCAM* is 2.9% when the *S* is elevated from 16 to 64. We also assessed CBIR tasks based on continuous Euclidean and binary Hamming codes. We observed that mAP performance in Hamming space is lower than in the Euclidean space due to information loss during the binarization process. On average, the increase in mAP performance of *OCAM* in Euclidean space is 2.8% compared to the Hamming space. Lastly, CIBR performance can also depend on the native imbalance between different classes of images in the repository. A general inspection of the results in Tables II-VII suggest that CBIR performance is higher in relatively more balanced KVASIR and X-RAY datasets compared to the imbalanced ISIC 2019 dataset.

TABLE V
RETRIEVAL PERFORMANCE IN KVASIR DATASET (HAMMING)

|  | S=16 | | | | | | S=64 | | | | | |
|---|---|---|---|---|---|---|---|---|---|---|---|---|
|  | P@5 | P@20 | P@50 | P@100 | P@150 | mAP | P@5 | P@20 | P@50 | P@100 | P@150 | mAP |
| *Neural Codes* [39] | 85.41 | 82.86 | 80.70 | 78.24 | 74.08 | 75.17 | 86.95 | 83.77 | 81.11 | 80.09 | 76.24 | 78.89 |
| *Contrastive* [10] | 80.88 | 77.19 | 75.64 | 73.11 | 71.72 | 72.96 | 88.80 | 86.13 | 85.29 | 83.66 | 79.51 | 81.23 |
| *Triplet* [15] | 86.42 | 83.68 | 81.84 | 79.35 | 73.47 | 75.07 | 89.53 | 87.75 | 87.21 | 86.11 | 82.25 | 84.47 |
| *TriEP* [17] | 84.10 | 82.93 | 80.85 | 78.38 | 72.99 | 72.74 | 83.93 | 81.06 | 79.78 | 78.13 | 74.57 | 71.03 |
| *WABT* [18] | 76.08 | 74.66 | 73.62 | 73.10 | 70.67 | 63.94 | 74.83 | 73.96 | 73.45 | 73.23 | 71.20 | 67.13 |
| *dmTri* [41] | 87.81 | 85.38 | 84.50 | 82.26 | 76.80 | 79.37 | 88.93 | 87.08 | 86.16 | 84.81 | 80.01 | 83.97 |
| *CondTri* [20] | 84.73 | 83.25 | 82.02 | 79.61 | 74.04 | 72.43 | 88.98 | 87.25 | 86.38 | 84.87 | 80.90 | 82.73 |
| *CTLL* [21] | 84.92 | 82.56 | 80.53 | 77.44 | 71.94 | 74.42 | 86.62 | 87.45 | 86.81 | 85.87 | 81.78 | 84.38 |
| *OCAM* | 89.10 | 87.28 | 86.33 | 84.95 | 79.32 | **83.22** | 90.83 | 89.22 | 88.59 | 87.66 | 83.68 | **85.93** |

TABLE VI
RETRIEVAL PERFORMANCE IN ISIC 2019 DATASET (HAMMING)

|  | S=16 | | | | | S=64 | | | | |
|---|---|---|---|---|---|---|---|---|---|---|
|  | P@5 | P@10 | P@20 | P@35 | mAP | P@5 | P@10 | P@20 | P@35 | mAP |
| *Neural Codes* [39] | 66.16 | 64.62 | 62.88 | 60.48 | 55.57 | 66.84 | 65.37 | 63.79 | 62.11 | 56.38 |
| *Contrastive* [10] | 66.27 | 64.78 | 64.09 | 62.55 | 54.74 | 69.36 | 66.91 | 65.77 | 64.48 | 58.11 |
| *Triplet* [15] | 74.74 | 72.57 | 70.88 | 69.97 | 64.15 | 77.99 | 73.36 | 71.96 | 71.91 | 67.29 |
| *TriEP* [17] | 71.50 | 71.15 | 69.21 | 68.16 | 62.12 | 75.72 | 73.03 | 71.22 | 70.08 | 64.14 |
| *WABT* [18] | 61.57 | 66.12 | 66.91 | 65.58 | 61.60 | 61.82 | 62.51 | 64.96 | 64.17 | 59.73 |
| *dmTri* [41] | 72.95 | 70.60 | 69.55 | 68.70 | 62.31 | 77.91 | 75.34 | 73.91 | 73.02 | 68.44 |
| *CondTri* [20] | 74.91 | 72.88 | 71.01 | 70.11 | 64.91 | 78.22 | 75.45 | 74.00 | 73.14 | 68.77 |
| *CTLL* [21] | 72.59 | 71.48 | 70.52 | 69.59 | 64.73 | 77.47 | 74.73 | 73.05 | 71.97 | 67.94 |
| *OCAM* | 75.63 | 73.93 | 72.12 | 71.36 | **65.98** | 78.94 | 76.33 | 74.87 | 74.21 | **71.48** |

TABLE VII
RETRIEVAL PERFORMANCE IN X-RAY DATASET (HAMMING)

|  | S=16 | | | | | | S=64 | | | | | |
|---|---|---|---|---|---|---|---|---|---|---|---|---|
|  | P@5 | P@20 | P@50 | P@100 | P@150 | mAP | P@5 | P@20 | P@50 | P@100 | P@150 | mAP |
| *Neural Codes* [39] | 85.06 | 83.54 | 82.72 | 82.18 | 79.43 | 79.07 | 86.89 | 84.38 | 84.28 | 83.62 | 81.27 | 79.25 |
| *Contrastive* [10] | 82.61 | 81.48 | 80.58 | 79.79 | 78.00 | 76.44 | 87.17 | 85.42 | 85.17 | 84.88 | 84.71 | 84.18 |
| *Triplet* [15] | 85.85 | 83.79 | 83.43 | 82.96 | 82.67 | 81.29 | 88.91 | 86.16 | 85.63 | 85.27 | 85.01 | 84.37 |
| *TriEP* [17] | 84.41 | 83.35 | 83.05 | 82.44 | 81.95 | 80.50 | 87.41 | 85.32 | 85.06 | 84.84 | 84.65 | 84.09 |
| *WABT* [18] | 87.28 | 85.89 | 85.70 | 85.30 | 84.82 | **83.11** | 87.89 | 85.58 | 85.22 | 85.01 | 84.89 | 84.40 |
| *dmTri* [41] | 86.10 | 83.93 | 82.97 | 82.45 | 82.22 | 80.43 | 88.55 | 86.11 | 85.49 | 85.21 | 85.00 | 83.90 |
| *CondTri* [20] | 85.76 | 83.99 | 83.45 | 82.81 | 82.28 | 79.55 | 88.01 | 85.54 | 84.91 | 84.62 | 84.40 | 83.38 |
| *CTLL* [21] | 86.17 | 84.12 | 83.44 | 82.88 | 82.41 | 80.98 | 88.64 | 86.19 | 85.67 | 85.39 | 85.06 | 84.16 |
| *OCAM* | 88.83 | 86.60 | 86.02 | 85.55 | 85.25 | 82.89 | 91.19 | 88.83 | 88.41 | 88.31 | 88.15 | **86.41** |

For systematic performance comparisons, here we focused on competing methods that shared a common representational learning framework to derive image embeddings of matching dimensionality. Several previous studies have reported CBIR results on some of the datasets examined here using traditional and deep-learning techniques. Traditional methods generally use hand-crafted approaches such as wavelet features [45], PHOG features [46], and bag-of-words [43]. Their retrieval performances often lag behind solutions involving a learning-based approach [47]. Recent learning-based studies have also considered improvements to retrieval performance via architectural modifications as opposed to learning objectives. Proposed frameworks include a CNN-based query-driven distance approach [42], a cauchy rotation invariance method (CRI-ResNet) with ResNet18 architecture [25], a DenseNet-121 architecture with random rotation [44] and an attention based CNN method [48]. Because our main focus in the current study was to evaluate learning objectives, we did not directly compare OCAM against these baselines. That said, a comparison of reported metrics suggests that *OCAM* might yield nearly 3.4% higher performance in KVASIR, 10.9% higher performance in ISIC-2019, and 4.9% higher performance in X-RAY. That said, it is likely that combining architectural improvements with the learning objective in OCAM might enable further performance benefits. It remains important for future work to systematically examine the relative benefits of architectural and learning objective contributions to retrieval performance.

Several lines of improvement might enable *OCAM* to further its performance in CBIR tasks. An inspection of retrieval performance with point-wise, pair-wise and triplet-wise methods considered here suggests that performance improves as a higher number of image samples are considered during representational learning. Thus, the quality of learned embedding vectors might be further improved by adopting advanced loss functions based on quadruplet learning [49]. Architectures that explicitly leverage self-attention mechanisms might enable the capture of more representative embedding vectors for images by better modeling spatial context [6, 50]. A task-agnostic approach to more sensitively capture the distributional properties of medical images might employ recent diffusion models [51, 52].

## VI. CONCLUSION

Triplet-wise methods for learning image embeddings promise superior performance over point- and pair-wise methods in CBIR tasks. However, the traditional triplet formulation can suffer from suboptimal segregation between positive and negative samples, and requires manual tuning of a margin value. Here, we introduced a new triplet-wise method, *OCAM*, that addresses these limitations for improved CBIR performance. *OCAM* was demonstrated on three medical datasets from divergent domains for CBIR tasks executed in Euclidean and Hamming spaces. Our results clearly indicate that *OCAM* outperforms state-of-the-art point-wise, pair-wise and triplet-wise learning approaches. Therefore, it holds great promise for automating query search in CBIR systems that aim to improve diagnostic accuracy while minimizing processing delays.